\documentclass{PoS}
\usepackage{amsmath}

\newcommand{\be}{\begin{equation}}
\newcommand{\ee}{\end{equation}}
\newcommand{\eq}[1]{eq.~\ref{#1}}
\newcommand{\fig}[1]{Fig.~\ref{#1}}

\newcommand{\Obar}{\overline{O}}
\newcommand{\tauint}{\tau_\mathrm{int}}
\newcommand{\tauintu}{\tau_\mathrm{int}^u}
\DeclareMathOperator{\var}{var}

\newcommand{\onecol}[2]{
  \begin{minipage}[t]{#1}{#2\vfill} \end{minipage}
}

\title{Autocorrelations in Hybrid Monte Carlo Simulations}

\ShortTitle{Autocorrelations in Hybrid Monte Carlo Simulations}

\author{  Stefan Schaefer$^a$ and \speaker{Francesco Virotta}$^b$\\
\llap{$^a$}Humboldt Universit\"{a}t zu Berlin, Institut f\"{u}r
Physik,  Newtonstr. 15, 12489 Berlin, Germany\\ CERN, Physics
Department, CH-1211 Geneva 23, Switzerland\\ \llap{$^b$}NIC/DESY,
Platanenallee 6, 15738 Zeuthen, Germany\\ E-mail:
\email{stefan.schaefer@cern.ch}, \email{Francesco.Virotta@desy.de}  }

\abstract{ Simulations of QCD suffer from severe critical slowing down
towards the continuum limit. This problem is known to be prominent in
the topological charge, however, all observables are affected to
various degree by these slow modes in the Monte Carlo evolution. We
investigate the slowing down in high statistics simulations and
propose a new error analysis method, which gives a realistic estimate
of the contribution of the slow modes to the errors.  \\
\vspace{0.0cm}
\hfill
\onecol{4.0cm}{\vspace{0.5cm}
  \it CERN-PH-TH/2010-278 \\ 
  DESY 10-210 \\ 
  SFB/CPP-10-115}
}

\FullConference{
  The XXVIII International Symposium on Lattice 
  Field Theory, Lattice2010\\
  June 14-19, 2010\\
  Villasimius, Italy
}

\begin{document}

\section{Introduction}

Reliable estimation of physical quantities in lattice QCD requires
that all possible sources of both systematic and statistical error are
kept under control. In the continuum limit $a\to0$ a severe  slowing
down of the topological charge has been observed
\cite{DelDebbio:2002xa}.  Such slowing down corresponds to an increase
in auto-correlation times beyond the naive scaling of the
algorithm. This behavior is expected to influence not only the
topological charge but also other observables and sets a limit on how
close to the continuum we can get.  However, given the need to
simulate at increasingly finer lattice spacings  in order to keep
cut-off effects under control, it is of general interest to think
about the consequences of analyzing data in presence of long
auto-correlation times.

Our study starts from the observation that in a simulation different
observables do have different auto-correlation times. The reason
behind this is related to the spectral structure of the transition
matrix that describes the stochastic evolution in Monte Carlo (MC)
time. This observation can be used to formulate a procedure that gives
conservative estimates  of the statistical errors. The method we
propose is about consistently using information from the slower sector
of the simulation to give a safer estimate of the error of the mean
value of observables that have shorter auto-correlation times.

\section{Auto-correlations of Markov Chains} 

The error analysis of lattice QCD data has to deal with the presence
of auto-correlations. This is a consequence of the fact that all known
simulation methods belong to the class of Markov chain Monte Carlo
(MCMC) algorithms.  In the following we present a brief overview of
the concepts and definitions used.  For a more detailed summary and
explanation we refer to \cite{Schaefer:2010hu} and references therein. 

 The auto-correlation function $\Gamma$ as a function of MC time $t$
is given by \be \Gamma_O (t) = \lim_{N\to\infty}\;
\frac{1}{N-t}\sum_{i=1}^{N-t} \,[\,O(q_{i+t})- \Obar\,]\,
[\,O(q_i)-\Obar\,]         \ee    where $N$ denotes the length of the
MC history and $\Obar$ is the mean value of the observable $O$ as
measured from the data. The integrated auto-correlation time is the
integral of the auto-correlation function  \be \label{tauInt}
\tauint(O)=\frac{1}{2}+\sum_{t=1}^{\infty}
\rho_O(t)\;,\quad\text{where} \quad\rho_O(t) =
\frac{\Gamma_O(t)}{\Gamma_O(0)}\;.  \ee  For a practical estimate, the
sum in \eq{tauInt} is in practice always truncated at some window $W$
(typically much smaller than $N$) whose size can be defined according
to a criterion which balances the statistical and systematic
uncertainties \cite{Wolff:2003sm}.  This naturally suffers from a truncation
bias that is asymptotically removed as we increase the length of the
total MC history and together move the truncation window towards
larger times. For short histories in which the autocorrelation
function itself is known with little precision, however, the
truncation bias can be sizable.  Since the integrated auto-correlation
time enters the error formula for $\Obar$ as    \be\label{error}
\delta\Obar=\sqrt{\frac{2\tauint(O)\var{[O]}}{N}}\;,    \ee     a better
estimator for $\tauint(O)$ would certainly lead to more reliable error
bounds on the average of observables of interest.

\begin{figure}[tb!]
\begin{center}
  \includegraphics{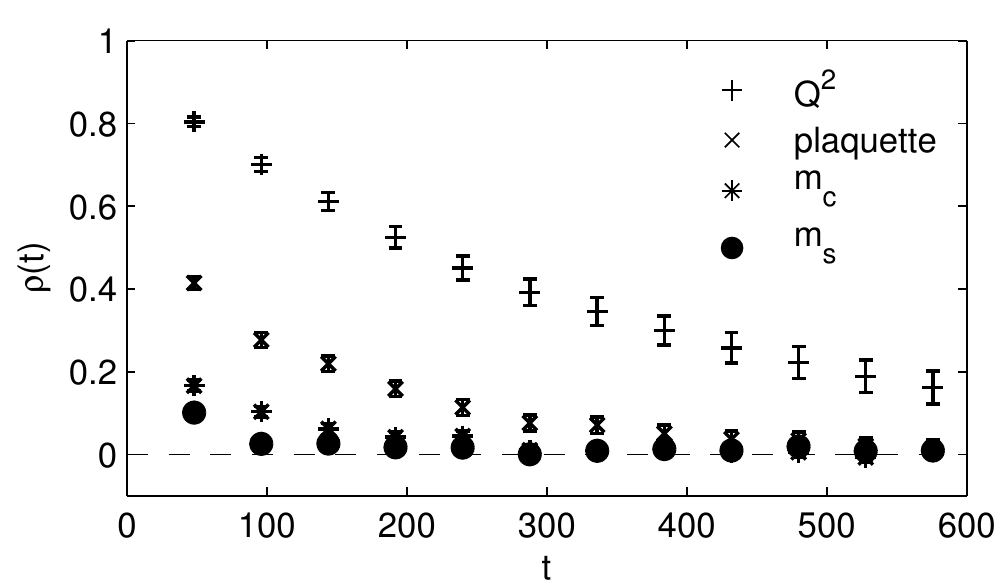}
\end{center}
\caption{\label{fig1} The normalized auto-correlation function for the
HYP smeared topological charge squared, the smeared plaquette and the
quenched pseudo-scalar mass of the $c\bar{c}'$ and of the $s\bar{s}'$
meson, computed respectively at the \emph{charm} and at the
\emph{strange} quark mass.  They are all computed at the same lattice
spacing of $a\approx0.074$fm. }
\end{figure}

By looking at the auto-correlation function of some  observables shown
in \fig{fig1}, all of them belonging to the same Markov chain, we can
immediately see how some of them decay much faster than others. From
the theory of Markov processes and some algorithmic considerations
(for example the assumption that the algorithm has detailed balance)
it is possible to derive the following spectral formula: \be 
\label{SpectralGamma} \Gamma_O(t) = \sum_{n\geq1}  (\lambda_n)^t\;
A_n(O) \,,\quad\text{where}\quad|\lambda_n|<1 \quad\text{and}\quad
A_n(O)\geq 0  \ee  where $\lambda_n$ are real eigenvalues of the
Markov matrix, ordered as
$\lambda_1\geq\lambda_2\geq\ldots\geq\lambda_n$ (again we refer to
\cite{Schaefer:2010hu} for details on the derivation). From \eq{SpectralGamma}
we can see that $\Gamma(t)$ is actually a linear combination of
decaying exponentials with positive coefficients.  An interpretation
of \fig{fig1} can then be given in terms of the amplitudes and the
time constants  $\tau_n = -1/\ln{\lambda_n}$ of the exponentials:
observables with auto-correlation functions that decay more slowly
have stronger coupling to the modes with larger time constant.

From a real world Monte Carlo simulation it is virtually impossible to
obtain a  definite knowledge about the longest time constants involved.
Since we need this information for the analysis, we therefore call
$\tau_*$ our best  estimate of the dominant time constant,
which we can either take from a model or by investigating a large
number of observables and take the largest observed value.  
Let us assume that for a given
observable $O$ all relevant time scales are smaller (or of the same
order) than the given $\tau_*$. If this is the case, we can choose a
window $W_u$ (best at a time where the auto-correlation function is
still significant), and define an  upper bound to the estimator of the
integrated auto-correlation time \be \label{upperTau}
\tauintu(O)=\frac{1}{2}+\sum_{t=0}^{W_u} \rho_O(t) + A_O(W_u)
\tau_*\;,\quad\text{with}\quad A_O(W_u)=\max\left(\rho_O(W_u)\,,\,
2\delta\rho_O(W_u)\right) \ee where $\delta\rho_O$ denotes the error
of the normalized auto-correlation function.

\section{A case study:  quenched pseudo-scalar meson mass} 

As a direct application of the ideas presented so far, we will examine
in detail the case of the auto-correlation function of a quenched
observable while critically applying to it the error formula in
\eq{upperTau}.  All the data we use is from a simulation with the
Wilson gauge action at lattice spacing $a=0.075$fm and volume of $(2.4
\text{fm})^4$. The algorithm used is the Domain Decomposed Hybrid
Monte Carlo (DD-HMC) \cite{Luscher:2005rx} reduced to the pure gauge
action, with a block size of $16\times8^3$. The size of our MC history
is of 145000 Molecular Dynamics units (MDU, that we will use
throughout the paper).  The observables that we consider are quenched
meson two-point functions, the plaquette and topological charge Q. In
order to be more sensitive to the slow modes, we also compute some
observables, in particular Q, on smoothed gauge fields. For this
purpose we apply  five levels of HYP smearing \cite{Hasenfratz:2001hp}
to the link variables.  

In order to apply \eq{upperTau} we first need a value of the time
constant $\tau_*$ used to compensate for truncating the tail
contribution due to the slow modes. Our proposal here is to define an
effective time constant $\tau_{\text{eff}}(t)$ by analogy with a QCD
effective mass \be \tau^{-1}_{\text{eff}}(t) = \log\left(
\frac{\Gamma(t)}{\Gamma(t+\delta t)}\right)/\delta t\;, \ee where
instead of computing it out of a two-point correlator we extract it
from the auto-correlation function of the observable that has the best
signal. The time  $\tau_*$ can then be  determined from a plateau
average in an ``effective mass'' plot.  This method will work well
with observables that strongly couple to possibly a single (or a few
very closely spaced) slow mode(s). 

\begin{figure}[tb!]
\begin{center}
  \includegraphics{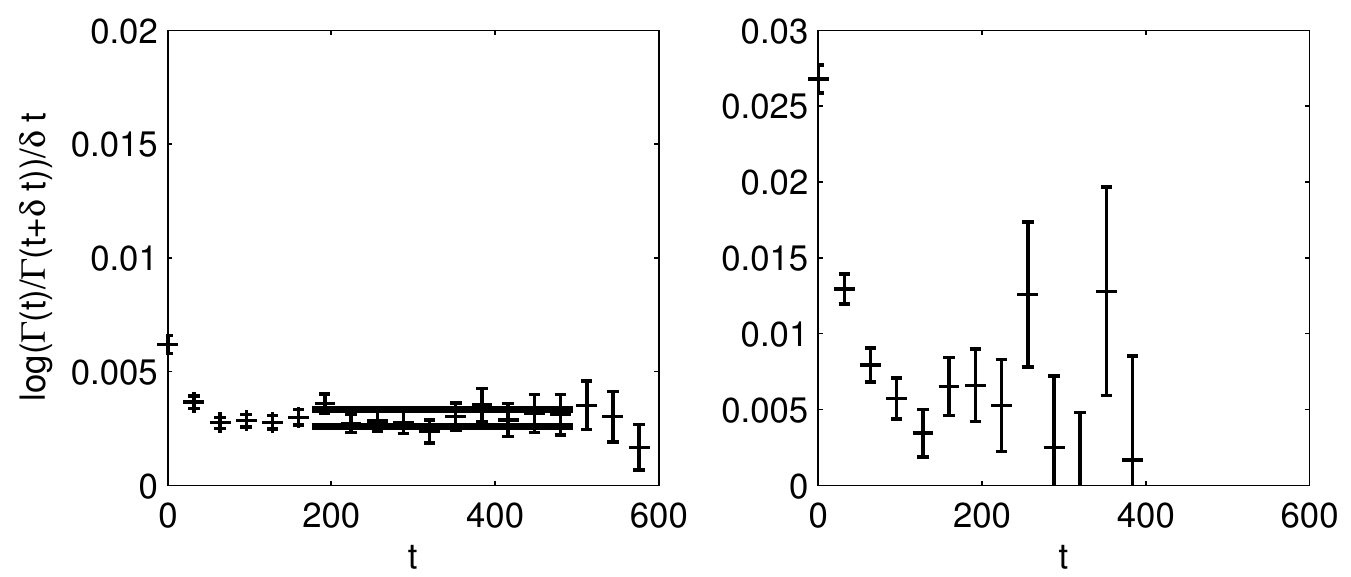}
\end{center}
\caption{\label{fig2} {\bf (Left)} ``Effective mass'' plot for the
auto-correlation function of the HYP smeared topological charge
squared. {\bf (Right)} Same plot for the smeared plaquette. }
\end{figure} 

The slowest observable that we consider is the parity even $Q^2$
instead of the parity odd $Q$. The reason behind this choice is that
even though in \eq{SpectralGamma} all modes contribute to $\Gamma$ for
a given observable, if the integrator preserves a symmetry of the
discretized action (in our case parity), it is possible to show that
contributions come only from algorithmic modes in the same sector as
the observable under study \cite{Schaefer:2010hu}.  As shown in \fig{fig2}
this method is giving a very good plateau for the topological charge
squared, while for the plaquette  results are not as
satisfactory. This has to do with the fact, already detectable in
\fig{fig1}, that, in this particular case, the smeared plaquette
couples   more strongly to modes with time constant(s) $\tau_n$
smaller than the time scales determining the slow decay of $Q^2$.
This does not mean that with $\tau_*$ we are identifying the  slowest
algorithmic mode (that would be $\tau_1$), but with it we are
confident to have identified the slowest relevant mode for the error
analysis of observables we are interested in.  We obtain $\tau_*$ by
averaging over the plateau shown in \fig{fig2} and the value we
measure is $\tau_*=330(45)$. This value can now be used to estimate
the upper bound of the error of another observable.

\begin{figure}[tb!]
\begin{center}
  \includegraphics{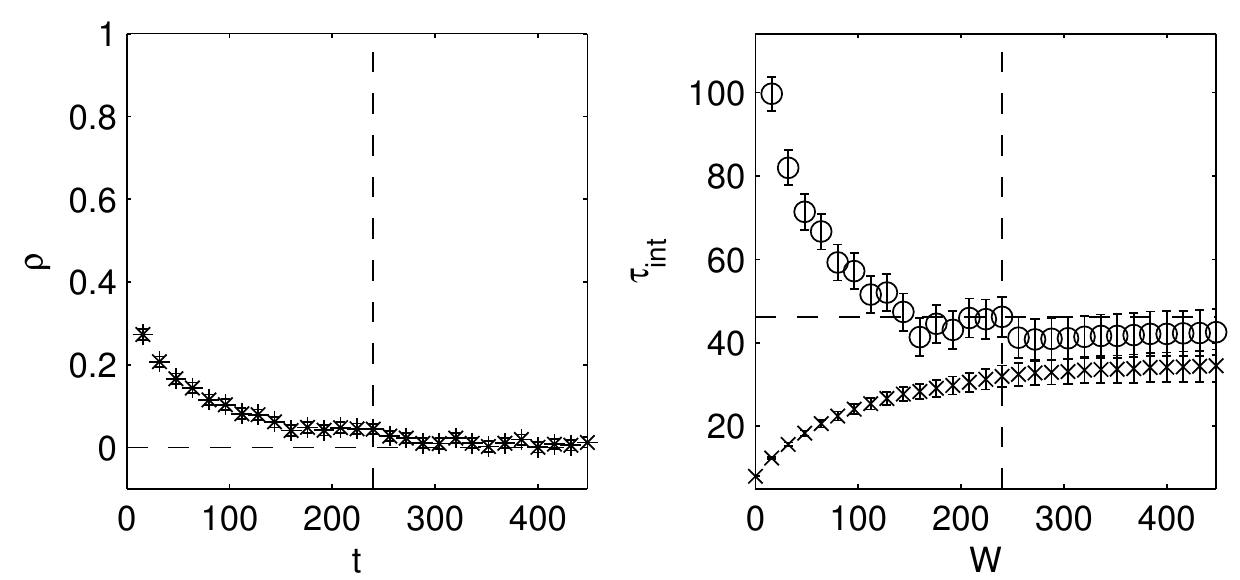}
\end{center}
\caption{\label{fig3} {\bf (Left)} Auto-correlation function of the
pseudo-scalar meson mass. The meson mass is obtained from a plateau
average. {\bf (Right)}  Sum of the normalized auto-correlation
function up to a window $W$. The cross indicates the truncated sum as
in \protect \eq{tauInt}. With the empty circle we denote the upper
bound as defined in \protect \eq{upperTau}. The vertical line denotes
the window chosen with the three sigma criterion, while the horizontal
marks the corresponding value of $\tauintu$.}
\end{figure}  For this second part of the analysis we will use the
mass of the quenched pseudo-scalar meson $c\bar{c}'$ (consisting of
two quarks with mass $m_c=m_{c'}=m_{\text{charm}}$) obtained by
averaging over a suitably chosen plateau.  The upper bound formula
\eq{upperTau} requires a prescription on how to choose the window
$W_u$. The criterion here used is that $\rho(W_u)\simeq 3\delta
\rho(W_u)$. If the auto-correlation function falls off very quickly,
then we choose the minimum $\tauintu$ between the one evaluated at
$W_u$ and the one evaluated at $W_u+\delta W$ (using \eq{upperTau}),
where $\delta W$ is the MC distance between measurements. In
\fig{fig3} we compare two estimators of $\tauint$: our  proposed upper
bound and the sum in \eq{tauInt} truncated at a window $W$ (without
compensating for the tail at $t>W$). At small window size it is
clearly visible that the upper bound overestimates the value of
$\tauint$, but at $W\sim100$ the estimator seems to settle on a
plateau that extends to much larger values of the window, eventually
overlapping with the ``lower bound'' estimates coming from below. This
is a first indication of the fact that in situations where a tail is
visible on the observable of interest (i.e. $m_c$) a knowledge coming
from the slower sectors of the simulation (i.e. $\tau_*$) can
effectively be used for a more reliable, possibly semi-automated,
error analysis.

When we compare our method to the one in \cite{Wolff:2003sm}, where an
optimal truncation window $W_{\text{opt}}$ is chosen by minimizing the
sum of statistical and systematic error, we obtain, in this particular
case, a window $W_{\text{opt}}=W_u$ and a value of
$\tauint=30(2)$. Our method gives $\tauintu=47(4)$, translating in an
increase of $\approx 25\%$ on the estimated statistical error (see
\eq{error}). When specifically studying properties of the algorithm,
we propose to use the two methods as upper/lower bound estimates of
the true value of $\tauint$.  In case one is interested in a reliable
estimate of the error of physical observables however, we strongly
suggest the use of $\tauintu$. What we have shown so far is evidence
that the upper bound gives a considerable, yet reasonable, increase in
the error bars.  In the next section we provide more evidence that the
method still works reasonably well in presence of shorter MC
histories.

\section{Upper bound with lower statistics}    As a complementary
check of the validity of the arguments presented so far, it is
interesting to study the upper bound formula in a case in which the
length of the MC history is short (of the order of $10\times\tau_*$,
for example).  The reason for this is that, in the end, we want to
apply this method of error analysis to simulations performed with
dynamical fermions, where the computational costs  can be very high.
This makes a detailed study as the one presented here  
prohibitively expensive.  Here our the total statistics
is around $500\times\tau_*$, making it possible to perform a
statistical study by splitting the whole history in bins of $3500$ MD
units. On each of these shorter histories we have then calculated the
auto-correlation function. To test the method we have assumed that
$\tau_*$ is known (either from a previous study or from a model as
proposed in \cite{Schaefer:2010hu}). In practice we have used the value of
$\tau_*$ obtained from the auto-correlation function of $Q^2$ on the
entire history.

\begin{figure}[tb!]
\begin{center}
  \includegraphics{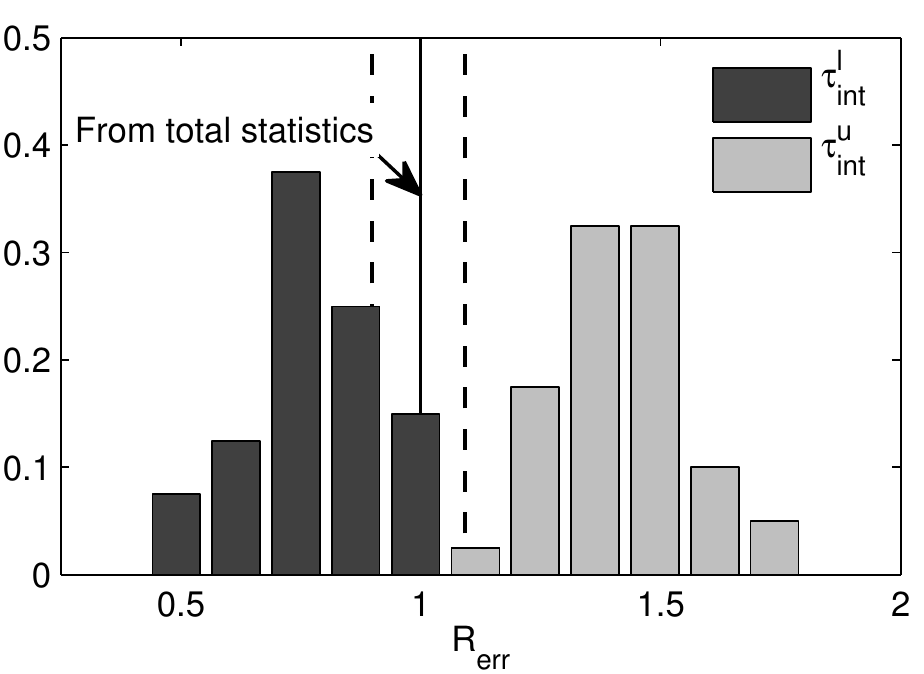}
\end{center}
\caption{\label{fig4}  Comparison between standard definition of
$\tauint^l$ and our improved estimate $\tauintu$, normalized by the
high statistics estimate, indicated by the vertical lines.
}
\end{figure} 

The comparison is shown in \fig{fig4}, in the form of two histograms
in which we have binned the values \be
R_\text{err}=\sqrt{\left(\tauint^{\text{Bound}}/\tauint\right)}\;, \ee
where $\tauint^{\text{Bound}}$ stands for either the upper bound
$\tauintu$ (computed with \eq{upperTau}) or the lower bound
$\tauint^l$ (computed by truncating the sum at $W_\text{opt}$, as
explained before) and $\tauint$ stands for the integrated
auto-correlation time obtained from the whole history (no splitting).
$R_\text{err}$ represents the relative deviation of the error
extracted from limited statistics to the more realistic error,
obtained from a more precise knowledge of the auto-correlation
time. We observe that with the standard method, the error is always
underestimated, up to a factor of two.  These distributions teach the
following lesson. The improved error estimate of \eq{upperTau} is
always safely close to the true error or somewhat above it.  An error
estimate using $\tauintu$ is recommended. The histograms also remind
us of an obvious fact: typically the error of the statistical error is
not that small in QCD simulations.

\section{Summary} In this study we have shown a method for safer error
estimates in the limit of low statistics/long auto-correlation
times. We first have applied the method to the error analysis of a
quenched observable, illustrating also a method for extracting the
contribution coming from the slowest known sectors of the simulation
(i.e. smeared topological charge square). 

We have then studied the method in the context of low statistics,
showing that also in this regime the resulting estimates are not
overly conservative and are therefore a safer way to determine the
error.  As discussed above, the proposed procedure for the improved
error estimate can be automated. For this purpose, we have written a
matlab routine that will be made publicly available soon.

\noindent \acknowledgments We would like to thank R. Sommer for the
fruitful collaboration on the subjects presented here and M. L\"uscher, F. Palombi 
and U. Wolff for many useful discussions. 
This work is supported by the Deutsche
Forschungsgemeinschaft in the SFB/TR~09 and by the European community
through EU Contract No.~MRTN-CT-2006-035482, ``FLAVIAnet''.  We thank
the John von Neumann institute for computing and the HLRN  for
allocating computer time for this project. Part of our runs  were
performed on the PAX cluster at DESY, Zeuthen.

\end{document}